\documentclass[pre,eqsecnum,showpacs,amssymb]{revtex4}
\usepackage{amsmath,graphicx}
\usepackage{epsfig}

\begin{document}

\title{Adsorption Kinetics of a Single Polymer on a Solid Plane}

\author{S. Bhattacharya$^1$, A. Milchev$^{1,2}$, V.G. Rostiashvili$^1$ , A.Y. Grosberg$^{1,3}$ and T.A. Vilgis$^1$}
\affiliation{$^1$ Max Planck Institute for Polymer Research
  10 Ackermannweg, 55128 Mainz, Germany\\
$^2$ Institute for Physical Chemistry, Bulgarian Academy of Science, 1113 Sofia, Bulgaria\\
$^3$ Department of Physics, University  of Minnesota, Minneapolis, MN 55455, USA}

\begin{abstract}
We study analytically and by means of an off-lattice bead-spring dynamic Monte
Carlo simulation model the adsorption kinetics of a single macromolecule on a
structureless flat substrate in the regime of strong physisorption. The
underlying notion of a ``stem-flower'' polymer conformation, and the related
mechanism of ``zipping'' during the adsorption process are shown to lead to a
Fokker-Planck equation with reflecting boundary conditions for the
time-dependent probability distribution function (PDF) of the number of adsorbed
monomers. The theoretical treatment predicts that the mean fraction of adsorbed
segments grows with time as a power law with a power of $(1+\nu)^{-1}$ where
$\nu\approx 3/5$ is the Flory exponent. The instantaneous distribution of train
lengths is predicted to follow an exponential relationship. The corresponding
PDFs for loops and tails are also derived. The complete solution for the
time-dependent PDF of the number of adsorbed monomers is obtained numerically
from the set of discrete coupled differential equations and shown to be in
perfect agreement with the Monte Carlo simulation results. In addition to
homopolymer adsorption, we study also regular multiblock copolymers and random
copolymers, and demonstrate that their adsorption kinetics may be considered
within the same theoretical model.

\end{abstract}


\maketitle

\section{Introduction}

The adsorption of polymers at equilibrium is fairly well understood from
theoretical
\cite{Fleer,deGennes1,deGennes2,deGennes3,deGennes4,deGennes5,Eisen}, computer
simulation \cite{Eisen,Sommer,Baschnagel,Milchev} and experimental
\cite{Takahashi,Cohen} points of view.  On the other hand, a great deal of work
exist on the polymer non-equilibrium adsorbtion of single polymer chains. One of
the earliest MC - simulation along this line was implemented for single chains
on the cubic lattice \cite{Konst}. The authors have tested a totally
irreversible adsorption model and a reversible one where a move resulting in the
desorption of a segment was assigned a relative weight $\exp (\chi_s)$ with
$\chi_s$ being the segmental adsorption energy in units $k_BT$. It was found
that at $\chi_s > 2 k_BT$ the fraction of segments in loops and trains start to
deviate from their equilibrium values, i.e. the process become irreversible.
Eventually at  $\chi_s \approx 10 k_BT$ the fraction of segments in loops and
tails merges the corresponding values for the totally irreversible model.

One of the important questions concerns the scaling of the adsorption time
$\tau_{\rm ads}$ with the length of the polymer chain $N$. Shaffer
\cite{Shaffer} has studied this problem using Monte Carlo simulatitions with the
bond fluctuation model (BFM) for strong sticking energy of $10 k_BT$. He found
that the deviation in the instantaneous fraction of adsorbed monomers from its
equilibrium value can be described by a simple exponential decay. During the
late stages, however, the relaxation function begins to deviate from an
exponential, and the relaxation slows down considerably. According to Shaffer,
this might be due to artefacts of the lattice model (cf.also ref.\cite{Lai}).
The main result of Shaffer \cite{Shaffer} is that $\tau_{\rm ads} \sim N^{1.58}$
(for comparatively short chains $N\leq 80$).

The same scaling has been found by Ponomarev {\it et al.}\cite{Ponomar} who also
used the BFM for $N\leq 100$. Except the energy gain (per segment) of
$\epsilon_s$, an activation barrier $\epsilon_b$ for a segment to  desorb was
introduced in this simulation, defining thus a "temperature"  $T_b \equiv k_B
T/\epsilon_b$. This sets a characteristic time for the passage of a segment
across  the barrier $\tau_b = \tau_0 \exp(1/T_b)$. Different adsorption dynamics
has then been found, depending on the ratio of $\tau_b$ and the Rouse time:
$\tau_b/\tau_R \simeq N^{-2\nu-1} \exp(1/T_b)$. The case $\tau_b/\tau_R \ll 1$
(at $T_b \geq 1$ ) corresponds to  {\it strong physisorption}. On the other hand
if the chain is relatively short and the barrier is high enough ($T_b$ is low
enough) then $\tau_b/\tau_R \gg 1$, which corresponds to {\it chemisorption}.
They argue that at $\tau_b/\tau_R \ll 1$ the adsorption follows a {\it zipping
mechanism} whereby the chain adsorbs predominantly by means of sequential,
consecutive attachment of monomers, a process that quickly erases existing
loops. In this case $\tau_{\rm ads} \sim N^{1.57}$ for a SAW chain in agreement
with Shaffer's results~\cite{Shaffer}. In the opposite limit (chemisorption),
the presence of a barrier enhances loop formation in the course of adsorption.
It was shown that even a modest local barrier discourages the tendency for
zipping and switches on a new mechanism involving loop formation. The scaling
law in that regime reads $\tau_{\rm ads} \sim N^{\alpha}$, where the exponent
$\alpha = 0.8\pm 0.2$.

The irreversible chemisorption from the dilute polymer solution has been
theoretically studied \cite{O'Shaughnessy_1,O'Shaughnessy_2} by making use of the
master equation (ME) method \cite{Van_Kampen} for the loops distribution
function. The authors argue that the process is dominated by {\it accelerated
zipping} when the sequential adsorption is disrupted by large loops formation.

For strong physisorption the {\it simple zipping} mechanism, as opposed to the
accelerated zipping, has been also recently considered by Descas, Sommer and
Blumen~\cite{Descas}. The authors \cite{Descas} used the BFM and suggested a
simple theoretical description of the corresponding adsorption dynamics based on
what they call a ``stem - corona'' model. This  leads to the scaling prediction,
$\tau_{\rm ads} \sim N^{1+\nu}$, which is in reasonably good agreement with the
simulation result.

In the present paper we study the case of strong physisorption by means of an
off-lattice dynamic MC method and theoretically by employing the ME-formalism.
This makes it possible to describe the adsorption dynamics not  only in terms of
the average fraction of adsorbed segments but  also  to include train and tail
distribution functions which furnish the main constituents of the dynamic
adsorption theory. Section \ref{theory} starts with the description of the
adsorption dynamics model  which shares many common features with the one
suggested by Descas {\em et al.}~\cite{Descas}. Then we use this model within
the ME-formalism to treat the time evolution of the distribution of adsorbed
monomers (as well as the distributions of the monomers forming trains and
tails). It is shown that the problem can be mapped onto a drift-diffusion
process governed by a Fokker-Planck equation. We obtain the numerical solution
of this equation and discuss its consequences. In Section \ref{model} we
briefly introduce the MC model. The MC-simulation results for  homopolymers as
well as for block- and random - copolymers in the regime of strong adsorption
are given in Section \ref{results}. We show that our MC-findings are in good
agreement with the theoretical predictions. We summarize our results
and conclusions in Section \ref{summary}. Some details of the train
distribution function calculation are relegated to the Appendix.

\section{Adsorption dynamics in terms of  train and tail distributions}
\label{theory}
Consider a single polymer molecule (grafted with one end to a flat structureless
surface) in an adsorption experiment which is repeated over and over again. The
monomer - surface interaction is considered attractive  with a sticking
energy $\epsilon = E_{1} - E_{2}$, where $E_{1}$ and $E_{2}$ are respectively
the monomer energies before and after the adsorption event.

\subsection{Stem-flower scenario: A macroscopic law}

\begin{figure}[bht]
\vspace{0.8cm}
\includegraphics[width=8cm, height=6cm]{z_n.eps}
\hspace{2cm}
\includegraphics[width=7cm, height=5cm]{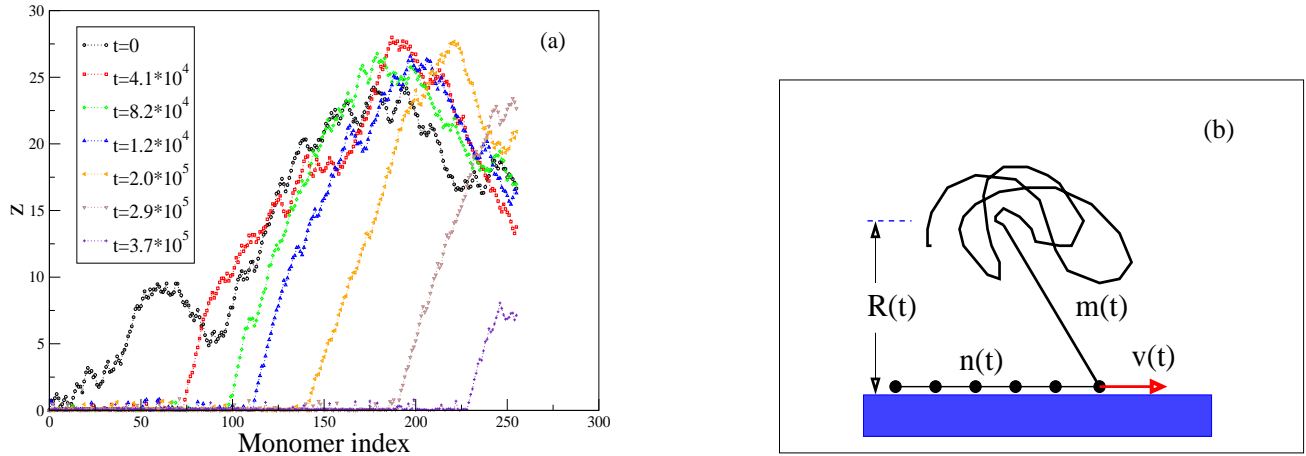}
\caption{ (a) Snapshots of an $N=256$ chain conformation, taken at successive
time moments during the adsorption process. The $z$-coordinate of the $i$-th
monomer  is plotted against monomer index $i$.  (b) Stem-flower picture of the
adsorption dynamics. The total number of adsorbed monomers at time $t$ is
denoted by $n(t)$. The tail which, contains all nonadsorbed monomers, consists
of a stretched part, a ``stem'', of length $m(t)$, and of a nonperturbed part
which is referred to as ``flower''. The rate of adsorption is $v(t)$. The
distance between the surface and the front of the tension propagation is $R(t)$.
\label{Stem_flower}}
\end{figure}

As indicated by earlier MC-simulation results~\cite{Ponomar,Descas}, in  the
strong physisorption regime the process is assumed to follow a simple zipping
mechanism. Figure ~\ref{Stem_flower}a gives snapshots of the chain conformation
as it is evident  from   our simulation results. One can see that the chain
conformation can be considered within the framework of a ``stem-flower'' picture
which was discussed first by Brochard-Wyart \cite{Brochard} as characteristic
for a polymer chain under strong stationary flow. Recently the ``stem-flower''
picture was employed in the case of non-stationary pulled polymer chain
\cite{Sakaure}. It should be pointed out that this picture shares many common
features with the "stem-corona" model, suggested by Descas {\it et
al}~\cite{Descas}. Here we reconsider it in a more systematic way and employ it
as a basic model to include fluctuations within the ME - formalism.

Fig.~\ref{Stem_flower}b presents schematically the stem-flower scenario of the
adsorption dynamics. The number of adsorbed monomers at time $t$ is denoted
by $n(t)$. The nonadsorbed fraction of the chain is subdivided into two parts:
a stretched part ("stem") of  length $m(t)$, and a remaining part  ("flower")
which is yet not affected by the tensile force of the substrate. The tensile
force propagation front is at distance $R(t)$ from the surface. The rate of
adsorbtion is denoted as $v(t)= a \frac{dn(t)}{dt}$, where $a$ is the chain
(Kuhn) segment length.

A single adsorption event occurs with energy gain $\epsilon$ and entropy loss
$\ln (\mu_3/\mu_2)$, where $\mu_3$ and $\mu_2$ are the connectivity constants in
three and two dimensions, respectively \cite{Vanderzande}. In result, the
driving force for adsorption can be expressed as
\begin{eqnarray}
f_{\rm drive} = \frac{\epsilon - k_BT \ln (\mu_3/\mu_2)}{a}= \frac{F}{a}
\label{Epsilon}
\end{eqnarray}
where $F = \epsilon - k_BT \ln (\mu_3/\mu_2)$ is the change in free energy. The
friction force is related to the pulling of the stem at rate $v(t)$, i.e.
\begin{eqnarray}
f_{\rm fric} = \zeta_0 \: a \: m(t) \: \frac{d n(t)}{d t}
\label{Friction}
\end{eqnarray}
where $\zeta_0$ is the Stokes friction coefficient of a single bead.
The equation of motion follows from the balance of driving, $f_{\rm drive}$,
and drag force, $f_{\rm fric}$, which yields
\begin{eqnarray}
\zeta_0 \: \: m(t) \: \frac{d n(t)}{d t} = \frac{F}{a^2}
\label{Eq_of_motion}
\end{eqnarray}

One may express $m(t)$ in terms of $n(t)$, if one assumes that at time
$t$ the  "flower" (which is placed on average at a distance $R(t)$ from the
surface) is not affected by the tensile force. This means that $R(t)$ is the
size which the chain portion $n(t) + m(t)$ occupied before the adsorption has
started, i.e.,
\begin{eqnarray}
a \left[ n(t) + m(t)\right]^{\nu} = R(t)
\label{Closure_1}
\end{eqnarray}
where $\nu$ is the Flory exponent (e.g., $\nu =3/5$ in $d=3$-dimensions)
\cite{Vanderzande}.
On the other hand, as shown in Fig.~\ref{Stem_flower}b,
\begin{eqnarray}
a \: m(t) \approx R(t)
\label{Closure_2}
\end{eqnarray}
up to a geometrical factor of order unity. Therefore the relation between $m(t)$
and $n(t)$ is given as
\begin{eqnarray}
n (t) \simeq m(t)^{1/\nu} - m(t)
\label{Closure_3}
\end{eqnarray}

During most of the adsorption process the stem is sufficiently long, $m(t) \gg
1$, so that $m(t)^{5/3}\gg m(t)$, i.e., $m(t) \simeq n(t)^{\nu}$ and
Eq.(\ref{Eq_of_motion}) becomes
\begin{eqnarray}
\zeta_0 \:  n(t)^{\nu} \: \frac{d n(t)}{d t} = \frac{F}{a^2}
\label{Eq_of_motion_2}
\end{eqnarray}
The solution of Eq. (\ref{Eq_of_motion_2}) reads
\begin{eqnarray}
n(t) \propto \left[ \frac{F}{a^2 \zeta_0} \: t\right]^{1/(1+\nu)}
\label{Solution}
\end{eqnarray}
In result, (for $d=3$ where $\nu = 3/5$) one obtains a law for the adsorption
kinetics, $n(t) \propto t^{0.62}$, which is in a good agreement with MC-findings
\cite{Shaffer,Ponomar,Descas}. In the course of adsorption the ``stem'' grows
and the ``flower'' moves farther away from the surface. This, as  it was
mentioned in ref.\cite{Descas}, makes the nucleation of a new adsorption site on
the surface less probable.

In the late stages of adsorption the "flower" has been largely consumed and
vanishes so that the non-adsorbed part of the macromolecule exists as a "stem"
only. From this moment on the closure relation reads
\begin{eqnarray}
n(t) + m(t) = N
\label{Later_stage_1}
\end{eqnarray}
Comparison of Eq.~(\ref{Later_stage_1}) with Eq.~(\ref{Closure_3}) shows that
this pure "stem" regime starts at $n(t) \geq N - N^{\nu} \approx N$, i.e.,
it could be basically neglected for sufficiently long chains.

The stem-flower scenario which we used in this section as well as the
macroscopic equation of motion, eq.(\ref{Eq_of_motion_2}), are employed below as
a starting point for the treatment of fluctuations.

\subsection{Time evolution of the distribution of adsorbed monomers}

Next we focus on the instantaneous number of adsorbed monomers (i.e., the total
train length) distribution function $P(n, t)$. The number of adsorbed monomers
 $n$ and the number monomers in the nonadsorbed chain tail $l$ are mutually
complementary, if one neglects the loops (we will argue below that in the strong
adsorption regime the loop contribution is rather small and reduces mainly to
loops of size unity). With this assumption, the corresponding tail
distribution function $T(l, t)$ reads
\begin{eqnarray}
T(l, t) = P(N-l, t).
\label{Tail_Distr}
\end{eqnarray}

\begin{figure}[bht]
\includegraphics[width=8cm, height=6cm]{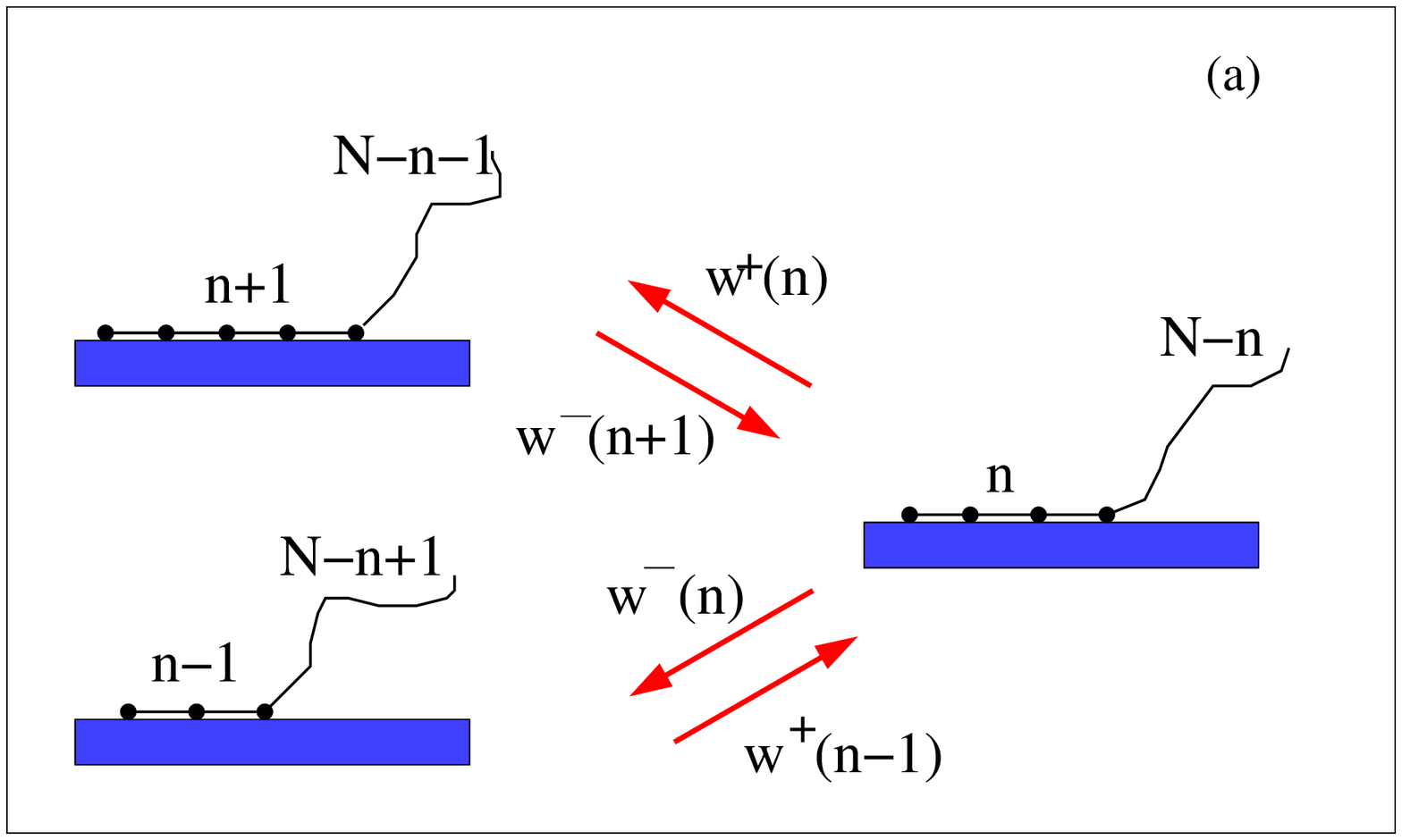}
\hspace{2.0cm}
\includegraphics[width=6cm, height=4cm]{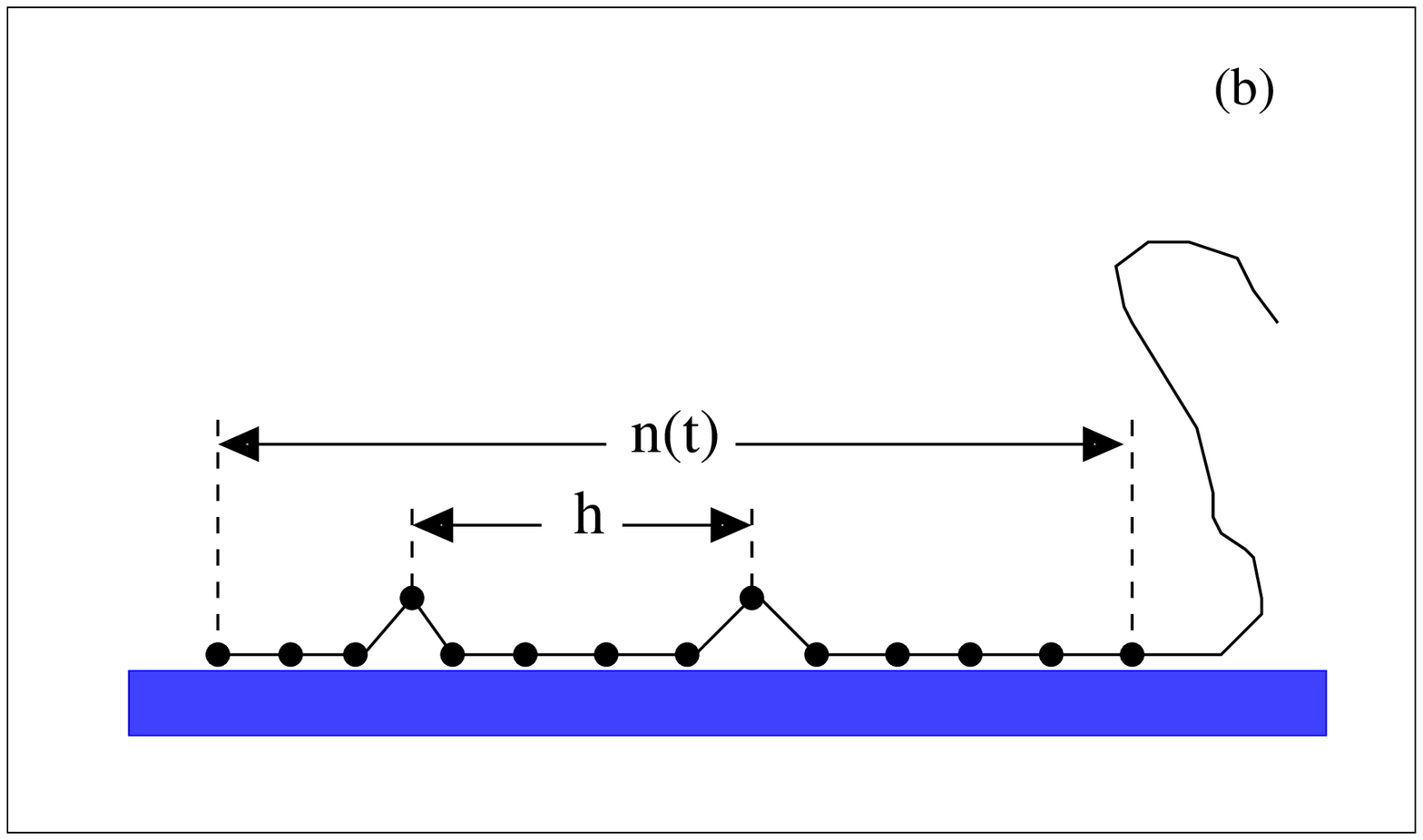}
\caption{(a) Creation - annihilation of an adsorption state with $n$-monomers
due to a single-step process. The arrows indicate possible single-step
transitions with $w^{+}(n)$ and $w^{-}(n)$ being the rate constants of
adsorption and desorption events, respectively. (b) The adsorbed monomers form
trains, divided by defects (loops of length unity). The total number of adsorbed
monomers at time $t$ is denoted by $n(t)$. The train length, $h$, itself is a
random number, subject to an exponential distribution $D(h,t)$ -
Eq.(\ref{Flory_Distr}).}
\label{One_step}
\end{figure}

Both $P(n, t)$ and $T(l, t)$ can be obtained either from the simulation or by
solving a set of coupled kinetic equations. For the latter we use the method of
the Master Equation~\cite{Van_Kampen}. We treat the adsorption as a sequence of
elementary events, describing the (un)zipping dynamics while keeping in mind
that within an elementary time interval only one monomer may change its state of
sorption. Thus one can treat the (un)zipping dynamics as an {\em one-step
process}, shown schematically in Fig.~\ref{One_step}a.

In order to specify the rate constants, we use the detailed balance
condition~\cite{Van_Kampen} which in our case (cf. Fig.~\ref{One_step}a) reads
\begin{eqnarray}
\frac{w^{+}(n-1)}{w^{-}(n)}
= {\rm e}^{F/k_BT}
\label{Detailed_balance_2}
\end{eqnarray}
where again $F = \epsilon - k_BT \ln (\mu_3/\mu_2)$ is the free energy win upon
a monomer adsorption event and the energy gain $\epsilon = E_1 - E_2$.
Detailed balance condition Eq.~(\ref{Detailed_balance_2}) is, of course, an
approximation for the non-equilibrium adsorption process in question. This
implies that, despite the global non-equilibrium, close to a ``touch-down''
point the monomers are in local equilibrium with respect to
adsorption-desorption events. This also means that the monomer size is small
enough as compared to the ``stem'' length, so that  this approximation is a good
one, compatible with the ``stem-flower''  picture of adsorption dynamics.

The detailed balance requirement fixes only the ratio of the rate constants and
does not fully determine their values which could be chosen as
\begin{eqnarray}
w^{-} (n) &=& q[m(n)] \: {\rm e}^{-F/k_BT}\nonumber\\
w^{+} (n-1)&=& q[m(n)] .
\label{W}
\end{eqnarray}
In Eq.~\ref{W} the {\em transmission} factor $q [m(n)]$ is determined by the
friction coefficient $\zeta$ which, within our stem-flower model, is defined as
$\zeta=\zeta_0 m$. Therefore, one obtains
\begin{eqnarray}
q[m(n)] = \frac{k_BT}{a^2 \zeta} = \frac{k_BT}{a^2 \zeta_0 \: m}.
\label{Q}
\end{eqnarray}
The notation $q[m(n)]$ implies that the stem length $m$ depends on the total
train length $n$ and, furthermore, the relationship $m(n)$ is given by the
closure Eq. (\ref{Closure_3}) which also holds for the instantaneous  values,
i.e.,
\begin{eqnarray}
n\simeq m^{1/\nu} - m
\label{Closure_4}
\end{eqnarray}

With the rate constants from Eq.~\ref{W} at hand, the one-step master equation
reads \cite{Van_Kampen}
\begin{eqnarray}
\frac{d}{d t} \:  P(n, t) &=& w^{-} (n+1)\: P (n+1, t) + w^{+} (n-1) \: P (n-1,
t)\nonumber\\
&-&   w^{+} (n) \: P (n, t) - w^{-}(n) \: P (n, t)
\label{One_Step_ME}
\end{eqnarray}
or, in a more compact form
\begin{eqnarray}
\frac{d}{d t} \:  P(n, t) = \Delta \left[ w^{-}(n) \: P (n, t)\right]  +
\Delta^{-1} \left[ w^{+}(n) \: P (n, t)\right]
\label{ME}
\end{eqnarray}
where the finite-difference operators $\Delta,\; \Delta^{-1}$ are defined as
\begin{eqnarray}
\Delta f (n) &\equiv& f (n+1) - f(n)\nonumber\\
\Delta^{-1} f (n) &\equiv& f (n-1) - f(n)
\label{Delta}
\end{eqnarray}

The total number of the adsorbed monomers varies between $1$ and $N$, i.e.,
$1\leq n \leq N$. For $n=1$ the  Eq.~(\ref{One_Step_ME}) has to be replaced by
\begin{eqnarray}
\frac{d}{d t} P(1, t) = w^{-}(2) P(2, t) - w^{+}(1) P(1, t)
\label{At_Left}
\end{eqnarray}
Similarly, for $n = N$ the ME reads
\begin{eqnarray}
\frac{d}{d t} P(N, t) = w^{+}(N-1) P(N-1, t) - w^{-}(N) P(N, t)
\label{At_Right}
\end{eqnarray}
Finally, the set of master equations (\ref{One_Step_ME}), (\ref{At_Left}) and
(\ref{At_Right}) should be supplemented by the initial condition
\begin{eqnarray}
P(n, t=0) = \delta (n-1)
\label{Initial_Cond}
\end{eqnarray}
because the adsorbtion starts from the state of a one chain end grafted at the
surface.

The equation of motion for the first statistical moment, $\left\langle
n\right\rangle = \sum_{n=1}^{\infty} n P(n, t)$, can be obtained from Eq.
(\ref{ME}) by performing the summation by parts:
\begin{eqnarray}
\sum_{n=0}^{N-1} \: g(n) \Delta \: f(n) = g(N) f(N) - g(0) f(0) + \sum_{n=1}^{N}
\: f(n) \Delta^{-1} g(n)
\label{Summation}
\end{eqnarray}
where $f(n)$ and $g(n)$ are arbitrary functions. Taking this into account and
keeping in mind that  $P(N, t)=P(0, t)=0$ for simplicity, the
equation of motion for $\left\langle n \right\rangle $ then yields
\begin{eqnarray}
\frac{d}{d t} \: \left\langle n\right\rangle = - \left\langle w^{-} (n)
\right\rangle  + \left\langle w^{+} (n) \right\rangle
\label{First_moment}
\end{eqnarray}
With the relations for the rate constants, Eqs.~(\ref{W}) and \ref{Q}, this
equation of motion becomes
\begin{eqnarray}
 \zeta_0 \: m(t)\: \frac{d}{d t} \:  n(t) = \frac{k_BT}{a^2} \: \left[ 1
- {\rm
e}^{-F/k_BT}\right]
\label{Eq_of_motion_3}
\end{eqnarray}
where for brevity we use the notations $n(t) = \left< n \right>$ and $m(t) =
\left< m \right>$.  The result, Eq.~(\ref{Eq_of_motion_3}), should be compared
with Eq.~(\ref{Eq_of_motion}) derived earlier by means of a simplified physical
consideration (see also paper \cite{Descas} where this result was obtained before us).  Formally,  Eq.~(\ref{Eq_of_motion_3}) transforms back into
Eq.~(\ref{Eq_of_motion}) when adsorption is very weak, $F/k_B T \ll 1$. 
Importantly, Eq.~(\ref{Eq_of_motion_3}) has the same structure as
Eq.~(\ref{Eq_of_motion})  even when the adsorption is not weak and the quantity
$F/k_BT$ is not small; the only difference between these equations is that the
effective force in Eq.~(\ref{Eq_of_motion_3})  has the form $\left(k_B T /a
\right) \left[1 - e^{-F/k_B T} \right]$ instead of just $F/a$.  This can be
understood by the analogy with the second virial coefficient of interaction
between the monomer and the surface.  Indeed, we know that the contribution to
the free energy of an imperfect gas due to pair collisions is proportional to
the second virial coefficient rather than just interaction energy; similarly in
the case of adsorption, the effective second virial coefficient is the quantity
that describes the effect of monomer attraction to the wall.  Thus, the zipping
as a strongly non-equilibrium process can not be treated quasi-statically by
making use of a simple ``force-ballance''. The inclusion of fluctuations by
employing the ME-formalism is important in order to obtain the correct result
for the driving force.

\subsubsection{Fokker - Planck equation and boundary conditions}

It is instructive to change now from the discrete representation, Eqs.
(\ref{ME}), (\ref{At_Left}) and (\ref{At_Right}),  to a continuous one, namely,
to the Fokker-Planck equation for the distribution function $P(n, t)$ with
proper boundary conditions. This can be done by the substitution
\begin{eqnarray}
\Delta &\simeq& \frac{\partial}{\partial n} + \frac{1}{2}
\frac{\partial^2}{\partial n^2} \nonumber\\
\Delta^{-1} &\simeq& -\frac{\partial}{\partial n} + \frac{1}{2}
\frac{\partial^2}{\partial n^2}
\label{Continuous}
\end{eqnarray}
After that, Eq.(\ref{ME}) takes on the form
\begin{eqnarray}
\frac{\partial}{\partial t} \: P(n, t)  = \frac{\partial}{\partial
n}\left\lbrace \left[ w^{-}(n) - w^{+}(n)\right] \: P(n, t)\right\rbrace
+\frac{1}{2} \frac{\partial^2}{\partial n^2}\left\lbrace \left[ w^{-}(n) +
w^{+}(n)\right] \: P(n, t)\right\rbrace
\label{Fokker_Planck}
\end{eqnarray}
where $\left[ w^{+}(n) - w^{-}(n)\right] $ and $\left[ w^{-}(n) +
w^{+}(n)\right]/2$ play the roles of drift velocity and diffusion coefficient,
respectively.

To  derive the proper boundary conditions we recall that at $n=1$ the ME has a
different form, given by Eq.(\ref{At_Left}). It is convenient to require that
Eq.~(\ref{One_Step_ME}) is  still valid with the additional condition
\begin{eqnarray}
\left[ w^{+}(n-1) P (n-1, t) - w^{-} (n) P (n, t)\right]_{n=1} = 0
\label{BC_Left}
\end{eqnarray}
i.e., the transitions between a fictitious state $n=0$ and the state $n=1$ are
also balanced.

Similarly, to reconcile  the equation at $n=N$, given by Eq. (\ref{At_Right}),
with the general ME, Eq. (\ref{One_Step_ME}), one should impose the condition
\begin{eqnarray}
\left[ w^{-}(n+1) P (n+1, t) - w^{+} (n) P (n, t)\right]_{n=N} = 0
\label{BC_Right}
\end{eqnarray}
which again expresses the balance between an artificial state $n=N+1$ and the
state $n=N$.

In order to gain a deeper insight into the boundary conditions given by
Eqs.(\ref{BC_Left}) and (\ref{BC_Right}) let us represent
Eq.~(\ref{One_Step_ME}) in the form
\begin{eqnarray}
\frac{d}{d t} P(n, t) = \Delta \left[ w^{-} (n) P(n, t) - w^{+} (n-1) P(n-1,
t)\right]
\label{Dfferent_Form}
\end{eqnarray}
This representation looks like a discrete version of the  continuity equation,
stating that the value in the square brackets is the probability current
(with a negative sign), i.e.,
\begin{eqnarray}
J (n) = w^{+} (n-1) P(n-1, t) - w^{-} (n) P(n, t)
\label{Current}
\end{eqnarray}
A comparison of Eq. (\ref{Current}) with Eqs. (\ref{BC_Left}) and
(\ref{BC_Right}) allows one to conclude that
\begin{eqnarray}
J(n=1)= 0 \qquad \mbox{and} \qquad J(n=N+1)= 0
\label{BC_Discrete}
\end{eqnarray}
i.e., one should impose {\it reflecting} boundary conditions on both ends of
the interval.

Within the Fokker-Planck formalism the probability current has the form
\begin{eqnarray}
J(n) = \left[  w^{+}(n) - w^{-} (n)\right] P (n, t) - \frac{1}{2} \:
\frac{\partial}{\partial n} \: \left\lbrace \left[  w^{+}(n) + w^{-} (n)\right]
P (n, t) \right\rbrace
\label{Current_2}
\end{eqnarray}

Thus the Fokker-Planck formalism makes it possible to map the  strong adsorption
case onto a one-dimensional random walk problem with drift and diffusion
coefficients given in terms of rate constants, Eq.~(\ref{Fokker_Planck}).
While such a description provides physical insight into the problem, from the
viewpoint of numerics it is much easier to deal with the ME discrete set Eqs.
(\ref{One_Step_ME}), (\ref{At_Left}) and (\ref{At_Right}). We will discuss  the
results of this solution in Sec. II D.

\subsection{Train distribution}

Our MC-simulation results show that the distribution of loops in case of strong
physisorption is mainly dominated by the shortest loops  of length unity. These
loops can be considered as defects during the process of zipping. Moreover, this
distribution sets on much faster than the time for complete adsorption. Thus one
may consider the total number of the adsorbed monomers $n(t)$ as a slow variable
in  comparison to the number of defects (or loops of length unity). The adsorbed
monomers can be seen as an array of trains, separated  by an equilibrium number
of defects (see Fig. \ref{One_step}b).
The partition function of this one-dimensional array can be determined
rigorously (see Appendix A). Thus, one derives an expression for the train
distribution function
\begin{eqnarray}
D (h, t) = \frac{1}{h_{\rm av}(t)} \: \exp\left[ - \frac{h}{h_{\rm
av}(t)}\right]
\label{Flory_Distr}
\end{eqnarray}
where $h_{\rm av}(t)$ is the average train length. Eq.(\ref{Flory_Distr}) is
nothing but the Flory-Schulz distribution which usually governs the molecular
weight distribution in equilibrium polymerization of a broad class
of systems, referred to as {\em living} polymers\cite{Greer}.

\subsection{Results from the ME numerical solution}

\begin{figure}[bht]
\includegraphics[width=8cm, height=6cm]{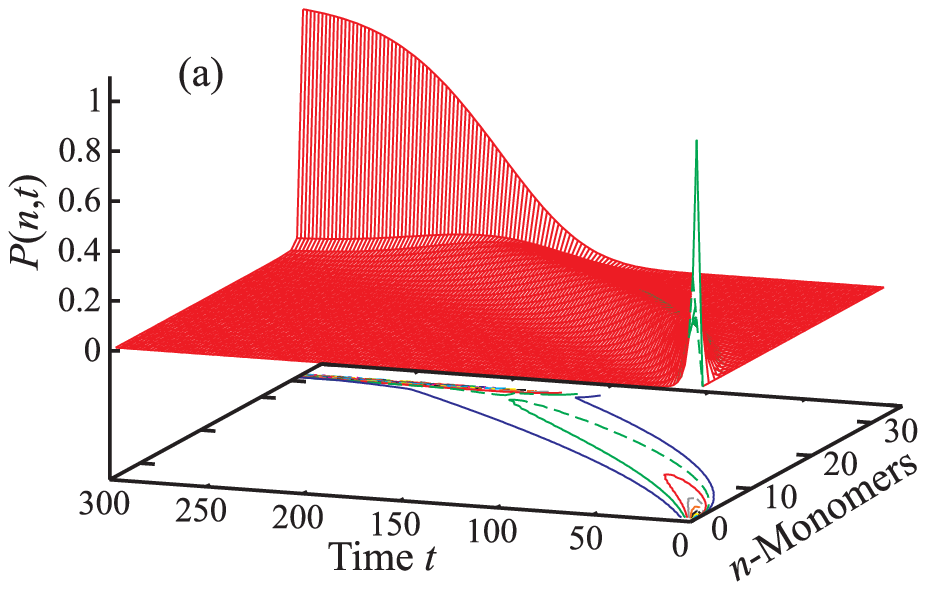}
\hspace{1.0cm}
\includegraphics[width=8cm, height=6cm]{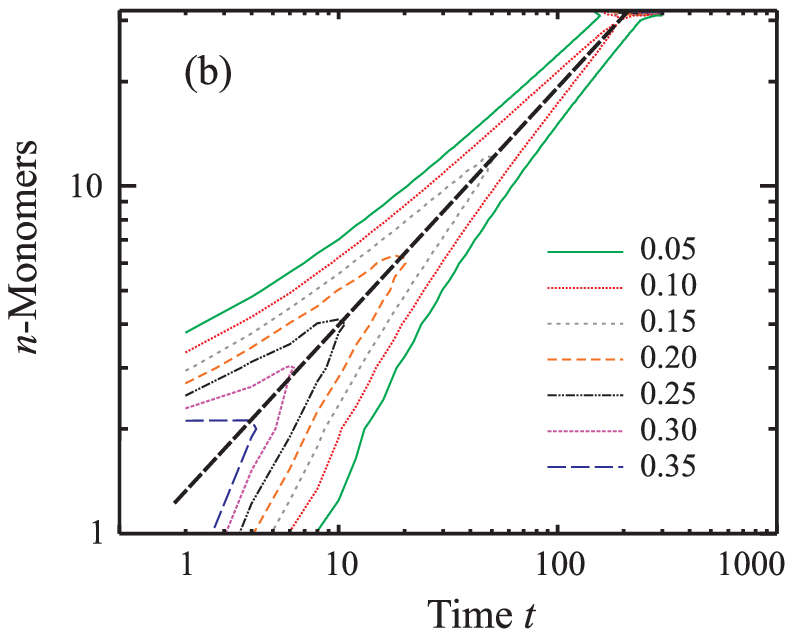} \caption{ Adsorbed monomer
number distribution function $P(n,t)$ (a) and its isolines as a $2$-dimensional
log-log plot (b). The variation of the distribution maximum, $n_{max}(t)$, is a
straight (dashed) line with slope $0.63$. \label{Plot3D}}
\end{figure}

The set of ordinary differential equations (\ref{One_Step_ME}), (\ref{At_Left})
and (\ref{At_Right}) with the initial condition, Eq.(\ref{Initial_Cond}), has
been solved numerically in this investigation. Typically, we use a chain length
$N=32$, the total time interval takes $300$ units of the the elementary time
$\tau_0=a^2 \zeta_0/k_BT$, the sticking energy was chosen (in units of
 $k_BT$) as $\epsilon =4.0$, whereas the entropy loss $\ln
(\mu_3/\mu_2) = \ln 2$. Figure \ref{Plot3D} demonstrates the result of this
solution.

\begin{figure}[bht]
\includegraphics[scale=0.4,angle=270]{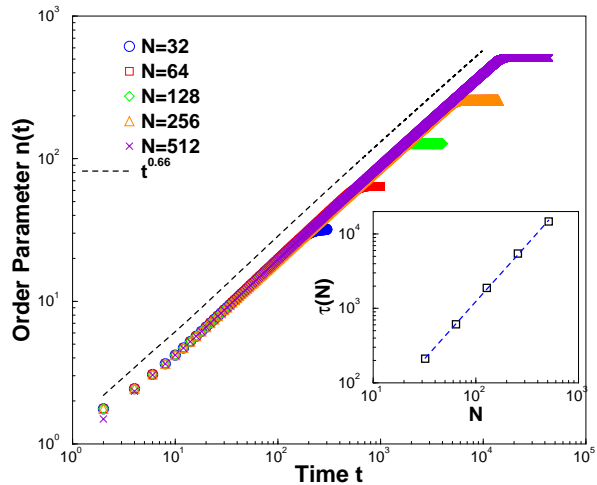}
\caption{The average adsorbed number of monomer vs. time for different chain
lengths $N$. Dashed line denotes the slope, $t^{0.66}$, following from
Eq.~\ref{Solution}. In the inset we show the resulting scaling of the
adsorption time with chain length, $\tau\propto N^{1.6}$. \label{MomentFirst}}
\end{figure}

As it can be seen from Fig.\ref{Plot3D}, the adsorbtion kinetics follows indeed
the drift - diffusion picture. The initial distribution is very narrow: the
adsorption starts with $n(0)=1$ as a grafted chain configuration. As time goes
by, the distribution maximum moves to larger adsorbed monomer numbers and the
distribution itself broadens. Eventually, the random process hits the boundary
$n = N$ and stays there due to  drift and the reflecting boundary conditions. As
 a result, the final distribution is a very narrow again, and is concentrated
around the boundary $n = N$.  It is of interest that in the double logarithmic
coordinates the distribution maximum  follows a straight line (cf. Fig.
\ref{Plot3D} right panel) which reveals a clear scaling law. The first moment
$n(t)$ of the distribution function $P(n, t)$ also exhibits well expressed
scaling behavior, $n(t) \sim t^{0.66}$,  as shown in Fig. \ref{MomentFirst}. In
the inset we also show the resulting relationship for the time of adsorption,
$\tau\propto N^{1.6}$, as expected from Eq.~\ref{Solution}. Based on the
numerical results for $P(n, t)$ and making use the relation,
Eq.~(\ref{Tail_Distr}), one can calculate the tail distribution function $T(l,
t)$ as well. We will discuss this in Sec. III where we present our MC-results.
There it will be seen that our MC-findings are in a good agreement with these
theoretical predictions.

\section{Monte Carlo Simulation Model}
\label{model}
To check the theoretical predictions mentioned in the previous sections
we have performed Monte Carlo simulations and investigated the adsorption
kinetics of a
homopolymer, multi-block copolymers, and random copolymers on flat surfaces.
We have used a coarse grained off-lattice bead spring model\cite{AMKB} to
describe the polymer chains. Our system consists of a single chain tethered at
one end to a flat structureless surface. There are two kinds of monomers: "A"
and "B", of which only the "A" type feels an attraction to the surface. The
surface interaction of the "A" type monomers is described by a square well
potential $U_w(z) =\epsilon $ for $z<\delta$ and $U_w(z) =0$ otherwise. Here
$\epsilon /k_BT$ is varied from $2.5$ to $10.0$. The effective bonded
interaction is described by the FENE (finitely extensible nonlinear elastic)
potential.
\begin{equation}
U_{FENE}= -K(1-l_0)^2\ln\left[1-\left(\frac{l-l_0}{l_{max}-l_0} \right)^2
\right]
\label{fene}
\end{equation}
with $K=20, l_{max}=1, l_0 =0.7, l_{min} =0.4$

The nonbonded interactions are described by the Morse potential.
\begin{equation}
\frac{U_M(r)}{\epsilon_M} =\exp(-2\alpha(r-r_{min}))-2\exp(-\alpha(r-r_{min}))
\end{equation}
with $\alpha =24,\; r_{min}=0.8,\; \epsilon_M/k_BT=1$.

We use periodic boundary conditions in the $x-y$ directions and impenetrable
walls in the $z$ direction. We have studied polymer chains of lengths $32$,
$64$, $128$, $256$ and $512$. Apart from homopolymers, we have also studied
copolymer chains with block size $M=1\div 16$ and random copolymers (with a
fraction of attractive monomers, $p=0.25,\;0.5,\;0.75$). The size of the box was
$64\times 64\times 64$ in all cases except for the $512$ chains where we used a
larger box size of $128\times 128\times 128$. The standard Metropolis algorithm
was employed to govern the moves with  self avoidance automatically incorporated
in the potentials. In each Monte Carlo update, a monomer was chosen at random
and a random displacement attempted with $\Delta x,\;\Delta y,\;\Delta z$ chosen
uniformly from the interval $-0.5\le \Delta x,\Delta y,\Delta z\le 0.5$. The
transition probability for the attempted move was calculated from the change
$\Delta U$ of the potential energies before and after the move as $W =
\exp(-\Delta U/k_BT)$. As for a standard Metropolis algorithm, the attempted
move was accepted if $W$ exceeds a random number uniformly distributed in the
interval $[0,1)$. A Monte Carlo Step (MCS) is elapsed when all $N$ monomers of
the chain are selected at random, and given the chance to perform an elementary
move. Before the surface adsorption potential is switched on, the polymer chain
is equilibrated by the MC method for a period of about $10^6$ MCS (depending on
the chain length $N$ this period is varied) whereupon one performs $200$
measurement runs, each of length $8\times 10^6$ MCS. In the case of random
copolymers, for a given composition, i.e., percentage $p$ of the $A-$monomers,
we create a new polymer chain in the beginning of the simulation run by means of
a randomly chosen sequence of segments. This chain is then sampled during the
course of the run, and replaced by a new sequence in the beginning of the next
run.

\section{Monte Carlo Simulation Results}
\label{results}

We present here the main results from the computer simulation of the adsorption
kinetics and compare them to those from the solution of the Master Equation,
Eqs.~(\ref{One_Step_ME}), (\ref{At_Left}) and (\ref{At_Right}), validating thus
the theoretical picture of Section~\ref{theory}.

\subsection{Order Parameter Kinetics - homopolymers}

\begin{figure}[htb]
\includegraphics[width=8cm, height=6cm]{OP_t.eps}
\hspace{1.0cm}
\includegraphics[width=8cm, height=6cm]{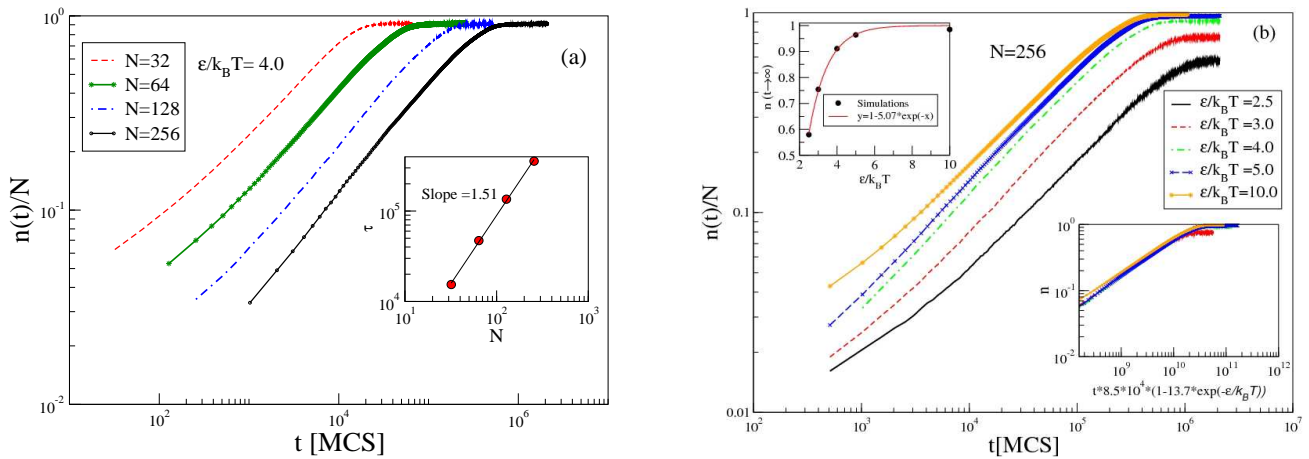}
\caption{(a) Time evolution of the order parameter (fraction of adsorbed
segments) for four different chain lengths $N=32,\;64,\;128,\;
\mbox{and}\; 256$ at surface potential $\epsilon/k_BT=4.0$. The slope of
the $N=256$-curve is $0.56$. The inset shows the scaling of the
adsorption time with chain length,$\tau\propto N^{1.51}$. The time $\tau$
is determined from the intersection point of the late time plateau with the
tangent $t^{0.56}$ to the respective $n(t)$-curve. (b) Adsorption kinetics for
different strengths $\epsilon$ of the surface potential. The variation of
the plateau height (i.e., the fraction of adsorbed monomers at equilibrium) with
$\epsilon$ is depicted in the upper inset where the solid line
$n_{t\rightarrow \infty}=1-5\exp\left( - \frac{\epsilon}{k_BT}\right )$
describes the equilibrium number of defects (vacancies). The lower inset shows a
collapse of the adsorption transients on a single 'master curve', if the time
axis is rescaled appropriately.\label{OP}}
\end{figure}

In Fig.~\ref{OP}a we show the adsorption time transients which describe the time
variation of the order parameter $n(t)/N$ (the fraction of adsorbed segments)
for homopolymer chains of different length $N$ and strong adhesion
$\epsilon/k_BT=4.0$. Evidently, in log-log coordinates these transients appear
as straight lines, suggesting that the time evolution of the adsorption process
is governed by a power law. As the chain length $N$ is increased, the slope of
the curves grows steadily, and for length $N=256$ it is equal to $\approx 0.56$.
This value is close to the theoretically expected slope of $(1+\nu)^{-1}\approx
0.62$ - cf. Eq.~\ref{Solution}, and for even longer lengths of the polymers
would most probably be observed. The total time $\tau$ it takes a polymer chain
to be fully adsorbed can be determined from the intersection of the respective
late time plateau of each transient with the straight line tangent to this
transient. Thus one may check the scaling of $\tau$ with polymer length $N$. In
the inset to Fig.~\ref{OP}a we show the observed scaling of the adsorption time
with chain length, $\tau\propto N^{\alpha}$ whereby the observed power ${\alpha
\approx 1.51}$ is again somewhat smaller than the expected one $1+\nu\approx
1.59$. This small discrepancy is most probably due to finite-size effects too.

Fig.~\ref{OP}b presents the adsorption transients for a chain of constant
length, $N=256$, for different strength of the surface potential.
Evidently, as the surface potential gets stronger, the final (equilibrium)
values of the transients at late times $t\rightarrow \infty$ grow while the
curves are horizontally shifted to shorter times. Notwithstanding, the slope of
the $n(t)$ curves remains unchanged when $\epsilon/k_BT$ is varied, suggesting
that the kinetics of the process is well described by the assumed zipping
mechanism.

\vskip 1.0cm
The changing plateau height may readily be understood as reflecting the
correction in the equilibrium fraction of adsorbed monomers due to the
presence of defects (vacancies) for any given value of $\epsilon/k_BT$. This is
demonstrated in the upper left inset in Fig.~\ref{OP}b where the observed
plateau values are shown to be perfectly described by the expression
$n_{t\rightarrow \infty}=1-5\exp\left( - \frac{\epsilon}{k_BT}\right )$
under the assumption that the probability of a monomer to desorb from the
surface (and create a vacancy in the train) is determined by the Boltzmann
factor $\exp\frac{-\epsilon}{k_BT}$. Evidently, the factor of $5$ in front of
the exponent yields the entropic gain in free energy when an adsorbed monomer
detaches from the surface while its nearest neighbors still stick to it.

The second inset in Fig.~\ref{OP}b shows that the adsorption time transients
collapse on a master curve, if one rescales the time axis appropriately. Note
that for a very strong potential, $\epsilon/K_BT=10.0$, the corresponding
transient deviates somewhat from the master curve since the establishment of
local equilibrium (which we assumed in the theory to happen much faster than
the adsorption process itself) is hampered. Also the transient for
$\epsilon/k_BT=2.5$ (not shown in this inset) was found not to fit into the
master curve since this strength is close to that of the critical
threshold for adsorption, the attraction to the surface is comparatively weak
and zipping is not the adequate  mechanism. For the transients which do
collapse on a master curve, however, one may view the rescaling of the time
axis in Fig.~\ref{OP}b by the expression $t\rightarrow
t [1 - 13.7\exp\frac{-\epsilon}{k_BT}]$ as a direct confirmation of
Eq.~\ref{Eq_of_motion_3} where the time variable $t$ may be rescaled
with the driving force of the process (i.e., with the expression in square
brackets). The factor $\approx 13.7$ gives then the ratio $\mu_3/\mu_2$ of the
effective coordination numbers in $3$- and $2$-dimensions of a polymer chain
with excluded volume interactions. $\mu_3$ and $\mu_2$ are model-dependent and
characterize, therefore, our off-lattice model.

\subsection{Order Parameter Kinetics - regular and random copolymers}

\begin{figure}[bht]
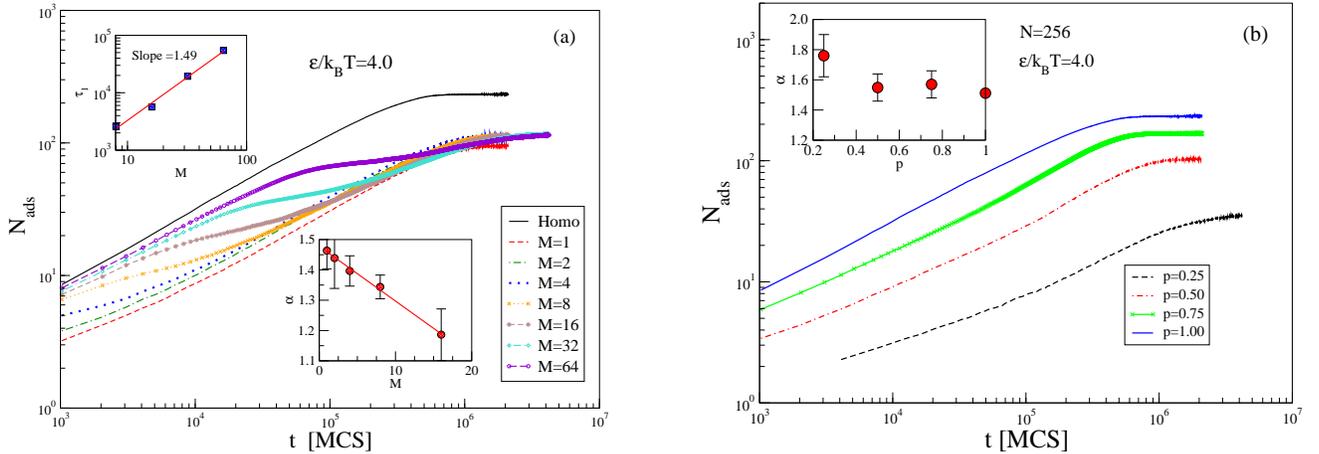

\includegraphics[width=8cm, height=6cm]{MBOP_t.eps}
\hspace{1.0cm}
\includegraphics[width=8cm, height=6cm]{OP_random.eps}
\caption{(a) Number of adsorbed segments, $N_{ads}(t))$, versus
time $t$ for regular $AB$-copolymers with block size $M=1\div 64$ and length
$N=256$. For comparison, the transient of a homopolymer is shown by
a solid line too. The time interval, taken by the initial
``shoulder'', is shown in the upper left inset. The lower inset
displays the variation of the scaling exponent, $\alpha$, for the
time of adsorption $\tau\propto N^\alpha$ versus block length
relationship. (b) The same as in (a) but for random copolymers of length
$N=256$ and different composition $p=0.25,\; 0.5,\; 0.75$. For $p=1$ one has
the case of a homopolymer. The inset shows the variation of
$\alpha$ with $p$.\label{MBOP}}
\end{figure}

In Fig.~\ref{MBOP} we examine the adsorption kinetics for the case of regular
block copolymers with block size $M$ - Fig.~\ref{MBOP}a, and for random
copolymers - Fig.~\ref{MBOP}b, bearing in mind that the zipping mechanism,
assumed in our theoretical treatment, is by no means self-evident when the file
of sticking $A$-monomers is interrupted by neutral $B$ segments. It becomes
evident from Fig.~\ref{MBOP}a, however, that, except for a characteristic
'shoulder' in the adsorption transients, the power-law character of the order
parameter variation with time remains unchanged. Evidently, only the first
shoulder in the adsorption transient is well expressed while the subsequent ones
disappear agaist the background of much larger time scales in the log-log
representation of Fig.~\ref{MBOP}a.  If, however, one monitors the adsorption of
only a {\em single} adsorption event with time then one observes in normal
coordinates a series of such shoulders like a 'staircase' in $N_{ads}(t)$ (not
shown here).

The variation of the power exponent, $\alpha$, with block length $M$,  where
$\alpha$ describes the scaling of the total adsorption time with polymer size
$N$, $\tau \propto N^\alpha$, is displayed in the inset right. Evidently,
$\alpha$ {\em declines} as the block size is increased. This finding appears
surprising at first sight, since it goes against the general trend of regular
multiblock copolymers resembling more and more homopolymers (with $\alpha =
1+\nu$ for the latter), as the block size $M\rightarrow \infty$. Moreover, it
would imply shorter adsorption times for smaller block size, $M \rightarrow
1$, although the shoulder length visibly grows with growing $M$ - see
Fig.~\ref{MBOP}a. In fact, however, as one may readily verify from
Fig.~\ref{MBOP}a, the transients are systematically shifted to longer times 
(i.e., the total adsorption takes longer) due to a growing prefactor for $M
\rightarrow 1$ which does not alter the scaling relationship $\tau \propto
N^\alpha$. One may thus conclude that the frequent disruption of the zipping
process for smaller blocks $M$ slows down the overall adsorption process (a
transient 'staircase' with numerous short steps) in comparison to chains with
larger $M$ where the zipping mechanism is fast (a 'staircase' with few longer
steps).
\vskip 1.0cm

The characteristic shoulder in the adsorption transients of regular multiblock
copolymers manifests itself in the early stage of adsorption and lasts
progressively longer when $M$ grows. We interpret the temporal length of this
shoulder with the time it takes for a segment from the {\em second} adsorptive
$A$-block in the polymer chain to be eventually captured by the attractive
surface, once the first $A$-block has been entirely adsorbed. For sufficiently
large blocks one would therefore expect that this time interval, $\tau_s$,
associated with the capture event, will scale as the Rouse time, $M^{1+2\nu}$,
of a non-adsorbing tethered chain of length $M$. The observed $\tau_s$ versus
$M$ relationship has been shown in the upper left inset in Fig.~\ref{MBOP}a. The
 slope of $\approx 1.49$ is less that the Rouse time scaling exponent, $2.18$,
which one may attribute to the rather small values of the block length $M$ that
were accessible in our simulation. One should also allow for scatter in the end
time of the shoulder due to the mismatch in the capture times of all the
successive $A$-blocks in the course of our statistical everaging over many
chains during the computer experiment.

In the case of random copolymers, Fig.~\ref{MBOP}b, the observed adsorption
transients resemble largely those of a homopolymer chain with the same number of
beads again, apart from the expected difference in the plateau height which is
determined by the equilibrium number of adsorbed monomers. One should note,
however, that a rescaling of the vertical axis with the fraction of sticking
monomers, $p$, does not lead to coinciding plateau height - evidently the loops
whose size also depends on $p$ also affect the equilibrium number of adsorbed
monomers. The variation of the observed scaling exponent $\alpha$ with
composition $p$ is shown in the inset to Fig.~\ref{MBOP}b wherefrom one gets
$\alpha \approx 1.6$. Note that this value is considerably lower than the power
of $2.24$ which has been observed earlier~\cite{Shaffer}, however, for very
short chains with only $10$ sticking beads. One may conclude that even for
random copolymer adsorption the typical time of the process scales as
$\tau\propto N^\alpha$, as observed for homo - and regular block copolymers. It
is conceivable, therefore, that an {\em effective}  zipping mechanism in terms
of renormalized segments, that is, segments consisting of an $A$ and $B$ diblock
unit of length $2M$ for regular multiblock copolymers provides an adequate
notion of the way the adsorption kinetics may be treated even in such more
complicated cases. For random copolymers the role of the block length $M$ would
then be played by the typical correlation length.

\subsection{Probability Distribution Functions}

\begin{figure}[htb]
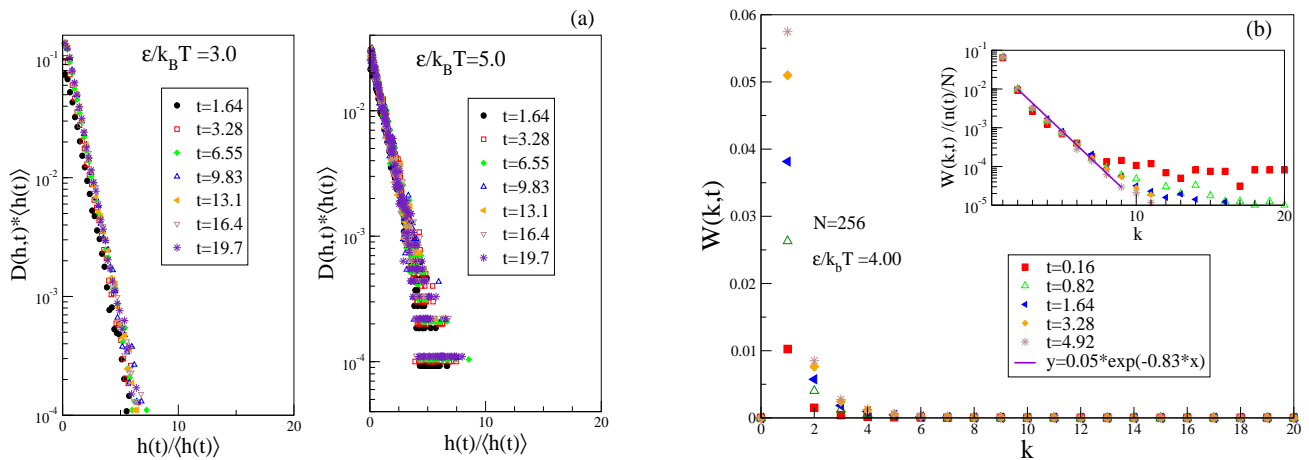

\includegraphics[width=8cm, height=6cm]{PDF_trains.eps}
\hspace{1.0cm}
\includegraphics[width=8cm, height=6cm]{PDF_loops.eps} \caption{(a)
Distribution of train lengths during the adsorption process of a homopolymer
chain with $N=256$ at two strengths of the adsorption potential $\epsilon$,
shown in semi-log coordinates. PDFs for different times (in units of $10^5$ MCS)
collapse on master curves when rescaled by the mean train length $h_{av}(t)$.
(b) Distribution of loop lengths $W(k,t)$ for $N=256$ and $\epsilon/k_BT=4.0$
during ongoing polymer adsorption. In the inset the PDF is normalized by $n(t)$
and shown to be a straight line in log-log coordinates.\label{PDF}}
\end{figure}

The time evolution in the corresponding Probability Distribution Functions (PDF)
of all the trains, loops and tails of adsorbed polymers provides a lot of
information and insight in the kinetics of the adsorption process. In the
Appendix we have derived theoretically the expected train distribution under the
assumption that local equilibrium of loops of unit length is established much
faster than the characteristic time of adsorption itself. The resulting
distribution of possible train lengths is shown to be exponential, in close
analogy to that of living polymers~\cite{Greer}. In Fig.~\ref{PDF}a we plot the
observed PDF of train lengths for a chain with $N=256$ at two strengths
$\epsilon/k_BT$ of the adsorption potential. When scaled with the mean train
length $h_{av}(t) = \langle h(t)\rangle$, at time $t$, in both cases for
$\epsilon/k_BT = 3.0\; \mbox{and} \;5.0$ one finds an almost perfect straight
line in semi-log coordinates, as predicted by Eq.~\ref{Flory_Distr}.
\vskip 1.0cm

One may thus conclude that the PDF for train lengths preserves its exponential
form during the course of the adsorption process, validating thus the
conjecture of rapid local equilibrium. The latter, however, is somewhat
violated for the case of rather strong adsorption - $\epsilon/k_BT=5.0$ -
shown in Fig.~\ref{PDF}a which is manifested by the increased scatter of data at
{\em late} times when the adsorption process overtakes to some extent the
relaxation kinetics on the surface. The PDF of loops $W(k,t)$ at different times
after the onset of adsorption is shown in Fig.~\ref{PDF}b. Evidently, the
distribution is sharply peaked at size one whereas less than the remaining
$20\%$ of the loops are of size two. Thus the loops can be viewed as single
thermally activated defects (vacancies) consisting of a desorbed single bead
with both of its nearest neighbors still attached to the adsorption plane. As
the inset in Fig.~\ref{PDF}b indicates, the PDF of loops is also described by an
exponential function. The PDFs for loops at different time collapse on a master
curve, if scaled appropriately with the instantaneous order parameter $n(t)/N$.
\vskip 1.0cm
\begin{figure}[bht]
\includegraphics[width=8cm, height=6cm]{tail_PDF_time.eps}
\hspace{1.0cm}
\includegraphics[width=8cm, height=6cm]{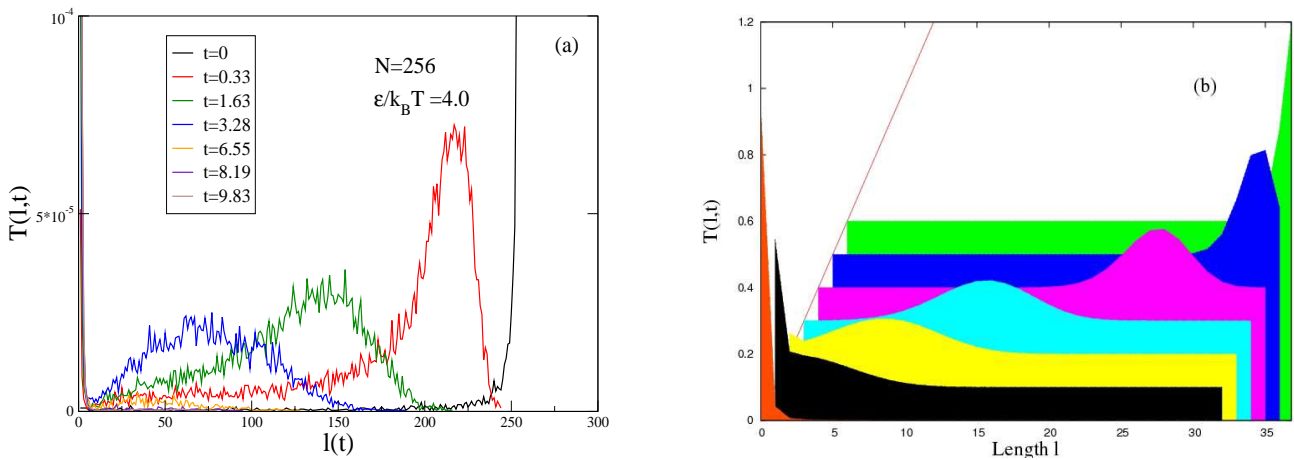}
\caption{(a) Distribution of tail size for different times (in units of $10^5$
MCS) during the polymer chain adsorption for a chain with $N=256$ at
$\epsilon/k_BT=4.0$. (b) The same as in (a) as derived from the solution of the
ME for chain length $N=32$. For better visibility the time slices for $t=1,\;
5,\; 30\;100,\; 150,\; 200,\; \mbox{and} \;300$ are shifted along the time axis
and arranged such that the initial distribution for $t=1$ is represented by the
most distant slice.\label{tail_time}}
\end{figure}

Eventually, in Fig.~\ref{tail_time}a we present the observed PDF of tails for
different times $t$ after the start of adsorption, and compare the simulation
results with those from the numeric solution for $T(l,t)$ according to
Eq.~\ref{Tail_Distr}. One may readily verify from  Fig.~\ref{tail_time} that the
similarity between simulational and theoretic results is really strong. In both
cases one starts at $t=1$ with a strongly peaked PDF at the full tail length
$l(t=1)=N$. As time goes by, the distribution  becomes broader and its maximum
shifts to smaller values. At late times the moving peak shrinks again and the
tail either vanishes, or reduces to a size of single segment which is expressed
by the sharp peak at the origin of the abscissa.

\section{Conclusions}\label{summary}

In this study we examine the adsorption kinetics of a single polymer chain on a
flat structureless plane in the strong physisorption regime. Adopting the
stem-flower model for a chain conformation during adsorption, and assuming the
segment attachment process to follow a ``zipping'' mechanism, we develop a
scaling theory which describes the time evolution of the fraction of adsorbed
monomers for polymer chains of arbitrary length $N$ at adsorption strength of
the surface $\epsilon/k_BT$.

We derive a Master Equation as well as the corresponding Fokker-Planck equation
for the time-dependent PDF of the  number of adsorbed monomers and for the
complementary PDF of tails, and define the appropriate reflecting boundary
conditions. Inherent in this derivation is the assumed condition of detailed
balance which makes it possible to relate the elementary steps of
adsorption/desorption. From the numeric solution of the equivalent discrete set
of coupled first-order differential equations we find that the growth of the
adsorbed fraction of monomers with time is governed by a power law, $n(t)\propto
t^{\frac{1}{1+\nu}}$, while the typical time of adsorption $\tau$ scales with
the length of the polymer $N$ as $\tau\propto N^\alpha$ with $\alpha = 1+\nu$.
The adsorption transients, found in the Monte Carlo simulation are in good
agreement with these predictions, if one takes into account the finite-size
effects due to the finite length of the studied polymer chains.

We demonstrate also that the height of the long time plateau in the adsorption
transients is determined by the equilibrium number of vacancies (defects) in
the trains of adsorbed monomers. The transients themselves are found to
collapse on a single master curve, if time is measured in reduced units which
scale with the corresponding driving force for adsorption as determined by the
surface potential $\epsilon/k_BT$.

A deeper insight into the adsorption kinetics is provided by our detailed study
of the relevant probability distributions of trains, loops and tails during the
adsorption. The predicted exponential expression for the PDF of trains is in
a very good  agreement with our simulational findings. The loops in the strong
physisorption regime are observed to reduce to occasional desorbed segments
(vacancies) which play little role in the dominating picture of trains and
tails. The PDFs of the latter are found from the simulation data to present a
shape which is fully consistent with that of the theoretic treatment. It should
be noted also that for chemisorption, a monomer adsorption event involves a
significant local activation barrier \cite{O'Shaughnessy_1,O'Shaughnessy_2}. In
this so-called ``accelerated zipping`` regime, the loops formation disrupts
the adsorption process and the corresponding dynamics differs significantly  from
the one investigated in this paper.

Eventually, in the case of regular multiblock and random copolymers we find
that the adsorption kinetics strongly resembles that of homopolymers. The
observed deviations from the latter suggest plausible interpretations in terms
of polymer dynamics, however, it is clear that additional investigations will
be warranted before a complete picture of the adsorption kinetics in this case
is established too.

\section{Acknowledgments}
We thank J.-U. Sommer for careful reading of the manuscript and helpful
discussion. A.~Milchev and A.~Grosberg are indebted to the Max-Planck Institute
for Polymer Research in Mainz, Germany, for hospitality during their visit in
the institute. A.~Grosberg acknowledges support from the Humboldt Foundation.
A.~Milchev and V.~Rostiashvili gratefully acknowledge support from the Deutsche
Forschungsgemeinschaft (DFG), grant No. SFB 625/B4.

\begin{appendix}

\section{Derivation of train distribution}
\label{Appendix}

The partition function of an one-dimensional array of $p+1$ trains, separated by
 $p$ defects, has the following form
\begin{eqnarray}
\Phi [n(t), p] &=& \mathop{\int \dots \int}\limits_{0<x_1<x_2\dots x_p<n(t)} d
x_1 \dots d x_p\nonumber\\
&=& \int\limits_{0}^{n(t)} \: dx_1 \int\limits_{x_1}^{n(t)} \: dx_2 \dots
\int\limits_{x_{p-1}}^{n(t)} \: dx_p = \frac{1}{p !} \: \left[ n(t)\right]^{p}
\label{Stat_sum}
\end{eqnarray}
where $n(t)$ is the total number of adsorbed monomers at time $t$.

Consider now the the distribution of an arbitrary train $h_{s+1} =
x_{s+1}-x_{s}$. In order to find it, one should carry out the integration in
Eq. (\ref{Stat_sum})
over all $x$-coordinates except $x_{s}$ and $x_{s+1}$. In result of the
integration one gets
\begin{eqnarray}
\Phi_{x_{s} x_{s+1}}[n(t), p] \:  d x_{s} d x_{s+1} = \frac{1}{(s-1)! (p-s-1)!}
\: x_{s}^{s-1} \: \left[n(t) - x_{s+1}\right]^{p-s-1} \: d x_{s} d x_{s+1}
\label{Stat_sum_2}
\end{eqnarray}
where Eq.(\ref{Stat_sum}) has been used separately for the intervals $[0,
x_{s}]$ and $[x_{s+1}, n(t)]$.

The distribution of the train length, $h_{s+1} = x_{s+1}-x_{s}$, follows
immediately from Eq.(\ref{Stat_sum_2}) after integrating over $x_{s}$, i.e.
\begin{eqnarray}
\Phi_{h_{s+1}}[n(t), p]  = \frac{1}{(s-1)! (p-s-1)!} \:\: \int\limits_{0}^{n(t) - h_{s+1}} \: x_{s}^{s-1} \: \left[n(t) - h_{s+1}-x_{s}\right]^{p-s-1} \: d x_{s}
\label{Stat_sum_3}
\end{eqnarray}
By the substitution, $y = x_{s}/[n(t) - h_{s+1}]$, in the integral of
eq.(\ref{Stat_sum_3}) one arrives at the result
\begin{eqnarray}
\Phi_{h_{s+1}}[n(t), p]  &=& \frac{\left[n(t) - h_{s+1}\right]^{p - 1} }{(s-1)! (p-s-1)!} \:\: \int\limits_{0}^{1} y^{s-1} (1 - y)^{p-s-1} \: d y \nonumber\\
&=& \frac{1}{(p - 1)!}\: \left[n(t) - h_{s+1}\right]^{p - 1}
\label{Stat_sum_4}
\end{eqnarray}
where one has used $\int_{0}^{1} y^{s-1} (1-y)^{p-s-1} d y = (s-1)!
(p-s-1)!/(p-1)!$.
The result in Eq. (\ref{Stat_sum_4}) does not depend on the consecutive number
of the train, as expected. The normalized probability to find a train of the
length $h$ at time $t$ is given by
\begin{eqnarray}
D (h, t) &=& \frac{\Phi_{h} [n(t), p]}{\Phi [n(t), p]} = \frac{p!}{(p-1)!} \:\frac{\left[ n(t) - h\right]^{p-1}}{\left[ n(t)\right]^{p}}\nonumber\\
&=& \frac{p}{n(t)} \: \left[ 1 - \frac{h}{n(t)}\right]^{p-1}\simeq \frac{p}{n(t)} \: \exp\left[ - h \frac{p}{n(t)}\right]
\end{eqnarray}
where one uses Eqs.(\ref{Stat_sum}) and (\ref{Stat_sum_4}) as well as the
conditions $p\gg 1$ and $h/n(t) \ll 1$. Taking into account that the average
train length $h_{\rm av}(t) = n(t)/p$, the last expression results in
Eq.(\ref{Flory_Distr}).

\end{appendix}

\end{document}